\documentclass[12pt]{article}
\usepackage{epsfig}
\usepackage{citesort}
\usepackage{times}

\newcommand{\re}[1]{(\ref{#1})}
\newcommand{\eq}{\begin{equation}}
\newcommand{\eqx}{\end{equation}}
\newcommand{\eqn}{\begin{eqnarray}}
\newcommand{\eqnx}{\end{eqnarray}}

\newcommand{\nin}{\noindent}

\newcommand{\el}{{\it e}}
\newcommand{\kphi}{\widetilde{\phi}}
\newcommand{\dk}{\Delta k}
\newcommand{\dv}{\Delta v}
\begin{document}
\nin

\begin{center}
{\Large \bf Inverse bremsstrahlung cross section estimated within evolving plasmas using effective ion potentials}\\
\vspace{10mm}

{\large F.~Wang $^{\ast}$, B.~Ziaja $^{\ast,\dag}$ \footnote{Author to whom correspondence should be addressed. Electronic mail: ziaja@mail.desy.de} 
and E.~Weckert $^{\ast}$}

\vspace{3mm}
$^{\ast}$ \it Hamburger Synchrotronstrahlungslabor, Deutsches Elektronen-Synchrotron, Notkestr. 85, D-22603 Hamburg, Germany\\

\vspace{3mm}
$^{\dag}$ \it Department of Theoretical Physics,
 Institute of Nuclear Physics,
\it Radzikowskiego 152, 31-342 Cracow, Poland\\
 
\end{center}

\vspace{10mm}
\centerline{\Large May 2008}

\vspace{6mm}
\nin
\vspace{6mm}
\nin

{\bf Abstract:}
{\em We estimate the total cross sections for field stimulated photoemissions and photoabsorptions by quasi-free electrons within a non-equilibrium plasma  evolving from the strong coupling to the weak coupling regime. Such transition may occur within laser-created plasmas, when the initially created plasma is cold but the heating of the plasma by the laser field is efficient. In particular, such a transition may occur within plasmas created by intense VUV radiation from a FEL as indicated by the results of the first experiments performed at the FLASH facility at DESY. 

In order to estimate the inverse bremsstrahlung cross sections, we use point-like and effective atomic potentials. For ions modelled as point-like charges, the total cross sections are strongly affected by the changing plasma environment. The maximal change of the cross sections may be of the order of $60$ at the change of the plasma parameters (inverse Debye length, $\kappa$, and the electron density, $\rho_e$), in the range: $\kappa=0-3$ \AA$^{-1}$ and $\rho_e=0.01-1$ \AA$^{-3}$. These ranges correspond to the physical conditions within the plasmas created during the first cluster experiments performed at the FLASH facility at DESY. In contrast, for the effective atomic potentials the total cross sections for photoemission and photoabsorption change only by a factor of 7 at most at the same plasma parameter range.

Our results show that the inverse bremsstrahlung cross section estimated with 
the effective atomic potentials is not much affected by the plasma environment. This observation validates the estimations of the enhanced heating effect obtained in \cite{santra,santra1}. This is important as this effect may be responsible for high energy absorption within clusters irradiated with VUV radiation. 
}\\ 


\section{Introduction}

Processes of photoabsorption and photoemission by quasi-free electrons within a plasma stimulated by an external laser field have been known and investigated for many years \cite{krain,krain1,kroll,pert,rozmus,santra,santra1,krenz}. The process of field stimulated absorption of radiation quanta is known as inverse bremsstrahlung (IB), and it is an inverse process to the stimulated photoemission. If the number of absorptions is larger than the number of emissions, the thermal energy of electrons increases with time. For a plasma in equilibrium the rates for these both processes obey detailed balance equations
\cite{meyer}.   

There exist various theoretical approaches to model the IB process (for review see \cite{rozmus,krenz}). Their applicability depends on the physical parameters of the system such as the drift velocity of electrons in the laser field, their thermal velocity, and the energy of radiation quanta. The applicability of some models is restricted only to a specific regime defined by these parameters \cite{rozmus}. In what follows we will use the quantum-mechanical approach to calculate the IB cross section, $\sigma_{IB}$, as proposed by Kroll and Watson in Ref.\ \cite{kroll}. This approach can be used to describe photoemission and photoabsorption by both slow and fast electrons. In particular, within this approximation, if the field strength parameter, $s={ {e{\bf E}_0}\over {\hbar \omega^2} }$ is small, and the free electrons are slow, single-photon exchanges dominate. The quantity, ${\bf E}_0$, denotes the electric field strength, $\omega$ is the photon frequency. For slow electrons, the photoabsorption of $n$ radiation quanta, $\gamma$, of energy $\hbar\omega$: $e(E_e)\pm n\gamma \rightarrow e(E_e\pm n\hbar \omega)$, may significantly increase the thermal energy of the electrons.

If $s$ is large, or if the free electrons are fast and undergo collisions with ions at large momentum transfers, multi-photon exchanges occur. This latter case can be described by the classical impact picture \cite{kroll} that is only valid if the drift component of the kinetic energy of the electron is much larger than the photon energy \cite{pert}. 

During heating the total energy absorption within the sample can be non-linear with respect to the exposure time and the pulse intensity: 
${{dE_{abs}}/{dt}} \propto N_{ion}(I,t)\,\sigma_{IB}(I)\,N_{el}(I,t)$,
as the total numbers of ions and electrons, $N_{ion}(I,t)$ and $N_{el}(I,t)$, change with pulse intensity and exposure time, and in addition the cross section, $\sigma_{IB}(I)$, is also a function of pulse intensity.

The IB process attracted much attention as a possible mechanism of efficient plasma heating when the results of the cluster experiments with vacuum ultraviolet (VUV) free-electron-laser (FEL) radiation performed at the FLASH facility at DESY became available \cite{desy,desy2,desy1,desy3,desy4,desy7}. These experiments covered the wavelength range from $100$ nm ($E_{\gamma}=12.7$ eV) down to $13$ nm ($E_{\gamma}=95.4$ eV). Pulse duration did not exceed $50$ fs, and the maximal pulse intensity was, $I\leq 10^{14}$ W/cm$^2$. 

In the first experiment performed at $100$ nm photon wavelength (VUV regime)
with $Xe_{2500}$ clusters highly charged Xe ions (up to $+8$) of high kinetic energies were detected. This indicated a strong energy absorption that could not be explained using standard theoretical approaches \cite{desy5,desy2,desy7}. More specifically, the energy absorbed was almost an order of magnitude larger than that one predicted with classical absorption models, and the ion charge states were much higher than those observed during the irradiation of isolated atoms at similar flux densities. This indicated that at this radiation wavelength some processes specific to many-body systems are responsible for the enhanced energy absorption. Several theoretical models have been proposed \cite{santra,siedschlag,bauer,brabec} which could explain various 
aspects of the increased photoabsorption and ionization dynamics observed 
in the experiments (for review see \cite{rost1}). The contribution of the IB process as a possible mechanism of the efficient electron heating was evaluated in detail in Refs. \cite{santra,santra1}. It was proposed that the strong energy absorption within an irradiated atomic cluster may result from the enhanced IB heating of quasi-free electrons. This enhanced IB rate was estimated with the effective atomic potential \cite{joa} which represents the attraction of the nucleus and the average screening effect of bound electrons surrounding the nucleus. Therefore, the distribution of electronic charge around the nucleus is smooth. An energetic electron that passes through the atom/ion is then scattered by the effective positive charge, $Z_{eff}$, larger than the net charge of the ion. This effect leads to an enhancement of the total IB rate that is proportional to the squared effective charge of the scatterer. This mechanism was first explored in Ref.\  \cite{santra}. Simulations of cluster irradiation
performed including this mechanism lead to the production of high charges 
within the irradiated clusters. These high charges were created in a sequence of electron impact ionizations. The ion charge state distributions were similar 
to those observed in the experiment \cite{desy}.

We stress here that the derivation of the IB rate from the effective atomic potentials as performed in Ref.\ \cite{santra,santra1} is in contrast to the standard approaches that assume Coulomb potentials of point-like ions \cite{krain1,krain,rozmus}. However, a heating mechanism similar to the one used in Ref. \cite{santra} was recently successfully tested in Ref. \cite{deiss}. It was applied to model the heating of quasi-free electrons in large rare-gas clusters irradiated with infrared laser pulses. These electrons were heated during elastic large-angle backscatterings on ion cores. The potentials of the ions were modelled with the parametrized atomic potential similar to that one in Ref. \cite{santra}. An absolute x-ray yield obtained with this effective atomic potential was in better agreement with the experimental data than that one obtained with the point-like atomic potential.

Here we aim to investigate in detail how the IB cross sections calculated using effective atomic potential from Refs. \cite{santra,deiss} depend on the changing plasma environment evolving from the strongly coupled to the weakly coupled regime. Up to our knowledge this question has not been addressed so far. Our results will validate the estimations of the heating rate obtained with the effective potentials in \cite{santra,santra1} by evaluating the impact of the changing plasma conditions on IB cross sections. 

As mentioned above, we will consider the limits of strongly and weakly coupled plasmas. The results obtained with effective atomic potentials will be compared to the IB cross sections calculated with point-like potentials. 

Our results will give estimates for the accuracy of the IB modelling within the evolving plasma in the regime relevant for the first cluster experiments \cite{desy}. Such estimates are important for performing the simulations of plasma heating, especially within plasmas that are created during the interaction of intense radiation with matter \cite{santra,santra1,deiss,ziajab,ziajab2,ziajab3}.

\section{Effective atomic potentials}

\begin{figure}
\vspace*{0.5cm}
\centerline{\epsfig{width=8cm, file=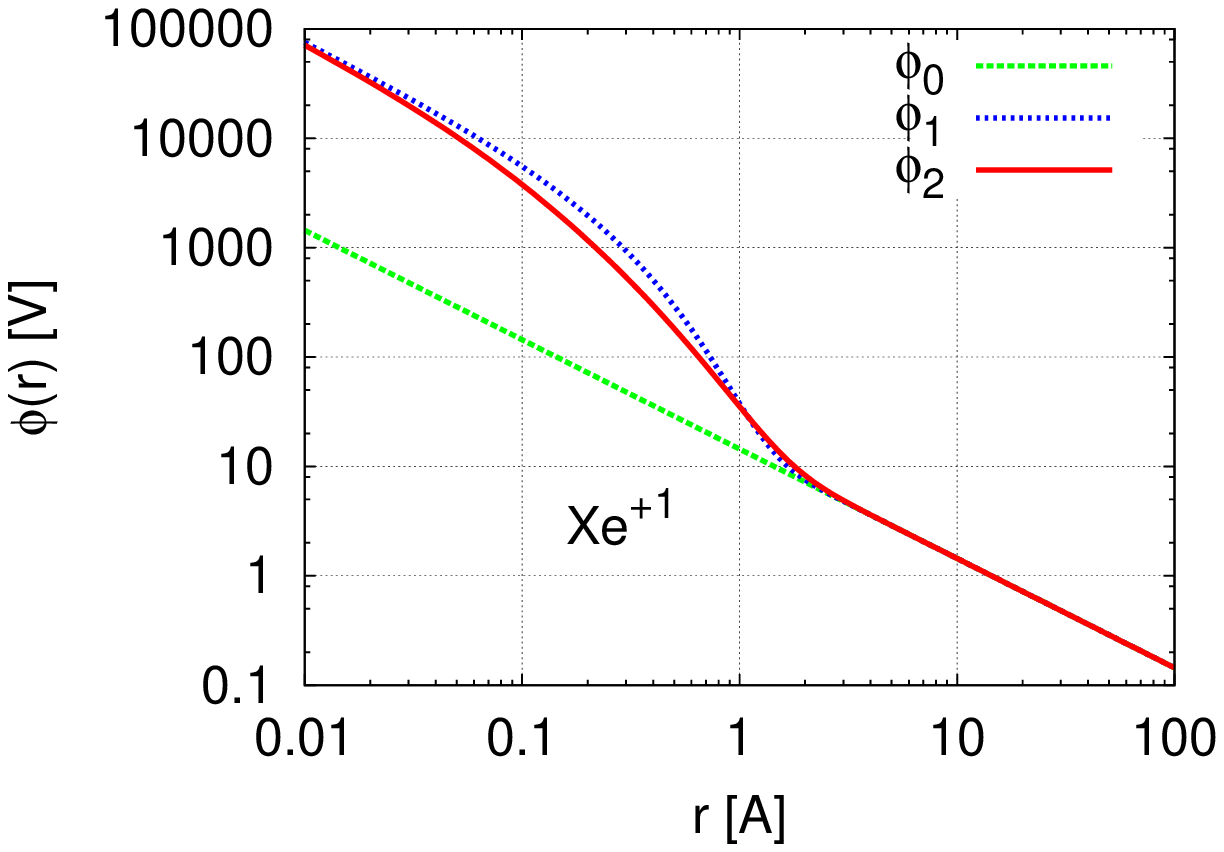}
\epsfig{width=8cm, file=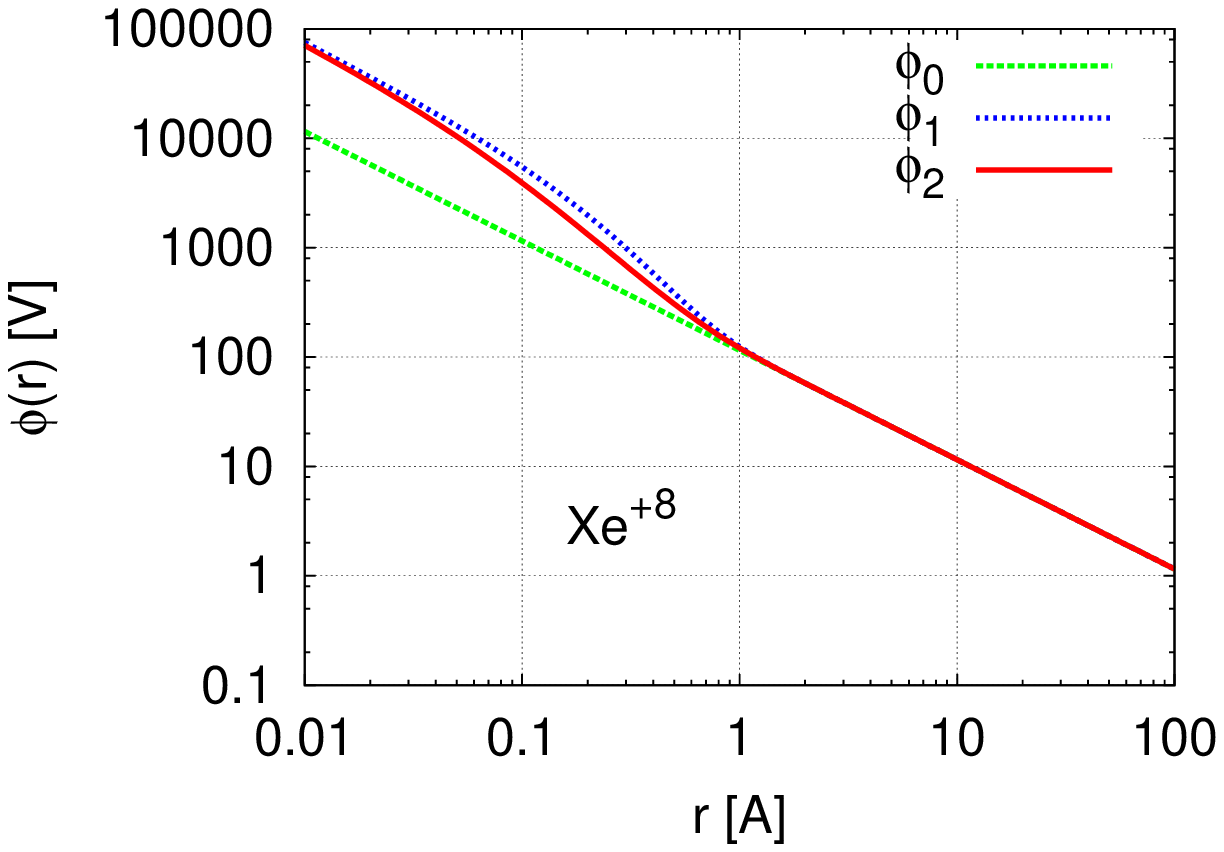}}

\caption{Effective atomic potentials for Xe ions. The potential applied in Ref. \cite{santra}, $\Phi_1(r)$, and the potential proposed in Ref. \cite{garvey}, $\Phi_2(r)$, are plotted for $Xe^{+1}$ ion (left), and $Xe^{+8}$ ion (right). The corresponding point-like potentials, $\Phi_0(r)$,  are also plotted for comparison.}
\label{potentials}
\end{figure}

For our tests we will use two different parametrizations of the 
effective atomic potential. The first one was also applied in Ref.\ \cite{santra} to describe the enhanced heating of electrons within atomic clusters irradiated with intense pulses of VUV radiation. The second one represents the independent-particle-model (IPM) potential introduced in Ref.\ \cite{garvey} and used in Ref.\ \cite{deiss} in order to estimate the cross section for elastic scattering of electrons on ions. The general form of these spherically symmetric potentials is:
\eq
\phi(r)={1\over\ 4\pi\epsilon_0 }\,\left( 
{{i \el}\over{r}}+{{(Z-i)\el}\over{r}}\,\Omega(r) \right),
\label{pot}
\eqx
where the charge $Z$ is the nuclear charge, $i=0,1,\ldots$ denotes the net ion charge. For point-like ions, $\phi_0(r)$, we have: $\Omega(r)=0$. 
Ref.\ \cite{santra} uses an exponential profile to model the screening by bound electrons: $\Omega_1(r)=e^{-\alpha_i r}$, where $\alpha_i$ is chosen so that the ionization energy of an ion calculated with this effective potential, $\phi_1(r)$, matches the corresponding experimental value. Ref.\ \cite{deiss} uses the independent-particle-model potential (IPM), $\phi_2(r)$, \cite{garvey}:
$\Omega_2(r)=\left[ {\eta \over \zeta}( e^{\zeta r}-1 ) +1 \right]^{-1}$, where
parameters $\eta$, $\zeta$ are element-specific and depend also on the ionization stage. We plot the different potentials for Xe ions at two ionization stages, $i=1$ and $i=8$ in Fig.\ \ref{potentials}. Despite the
different parametrization, the effective potentials are close to each other.
As expected, for small values of $r\leq 1$ \AA$\,$ (atomic size) there is a large discrepancy between the point-like and the effective atomic potentials. For larger values of $r$ the effective potentials approach the point-like potential.

The limiting values of $\Omega(r)$: $0<\Omega(r)<1$, are identical for both potentials and correspond to the physical limits of: i) the potential created by a pure nuclear charge, $Z$, at $r=0$, and ii) the potential created by net ion charge, $i$, at $r=\infty$.

The charge density, $\rho(r)$, that generates these effective potentials, is spherically symmetric and consists of the point-like positive nucleus charge, $Z$, screened by the cloud of bound electrons:
\eq
\rho(r)= 
{{Z \delta(r) }\over{4\pi r^2}}-{{(Z-i)}\over{4 \pi r}}\,\Omega^{''}(r),
\label{gest}
\eqx
where $\Omega^{''}(r)$ is the second derivative of $\Omega(r)$.
For a point-like ion of net charge, $i$, the corresponding charge density is:
\eq
\rho(r)={{i \delta(r) }\over{4\pi r^2}}
\eqx
%

\section{Weakly and strongly coupled plasma}

The temperature, density and charges of species determine the physical properties of plasmas. Two parameters are introduced in order to
classify various plasma regimes \cite{murillo}. The Coulomb coupling parameter, 
$\Gamma$, is defined separately for each plasma component as the ratio of its average potential energy to the average kinetic energy, $\Gamma\sim\mid \phi(r)/kT \mid$. If $\Gamma \gg 1$, plasma enters the strong coupling regime, where many-body screening effects are  significant. If $\Gamma \ll 1$, a plasma is considered to be ideal. The second parameter is the degeneracy parameter, $Y$, that is the ratio of the Fermi energy of a given plasma component to its average kinetic energy, 
$Y \sim \mid E_F/kT \mid$. For $Y \gg 1$ quantum statistics should be used (non-classical plasma). For $Y \ll 1$ the plasma can be treated classically.

\begin{figure}
\vspace*{0.5cm}
\centerline{\epsfig{width=10cm, file=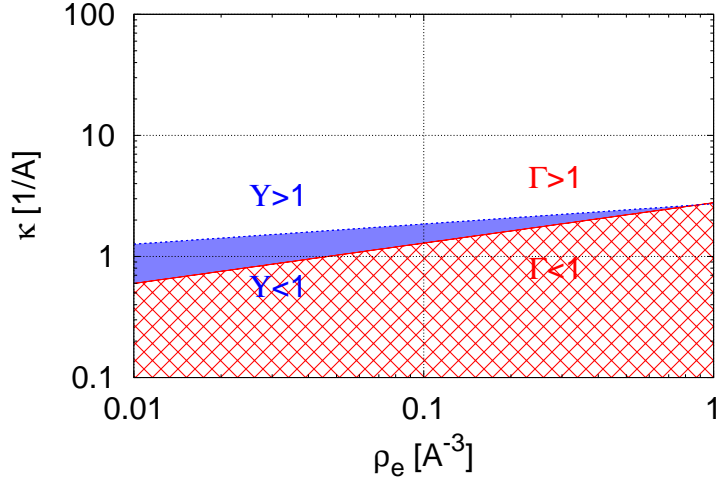}
}

\caption{Regions of strongly and weakly coupled plasma. Plasma parameters, $\Gamma$ and $Y$ are plotted as a function of screening parameter, $\kappa$ and electron density, $\rho_e$. Area filled with pattern corresponds to the regime of classical ideal plasma. Area filled with colour represents the regime of strongly coupled classical plasma. The remaining area represents the degenerate, strongly coupled plasma.}
\label{param}
\end{figure}

Various plasma regimes are plotted in Fig. \ref{param}. They correspond to the
estimated physical conditions within the plasmas created during the first cluster experiments performed at the FLASH facility at DESY \cite{desy}. For the typical electron densities within xenon clusters in the range of $\rho_e=0.01,0.1,1$ \AA$^{-3}$ plasmas enter the degenerate strong coupling regime if the Debye screening parameter, $\kappa\equiv\lambda_D^{-1}$ is  $> 1-3$ \AA$^{-1}$. In this regime the ion-sphere screening should be applied. For lower values of $\kappa$ the weak coupling approach is valid.

\begin{figure}
\vspace*{0.5cm}
\centerline{\epsfig{width=8cm, file=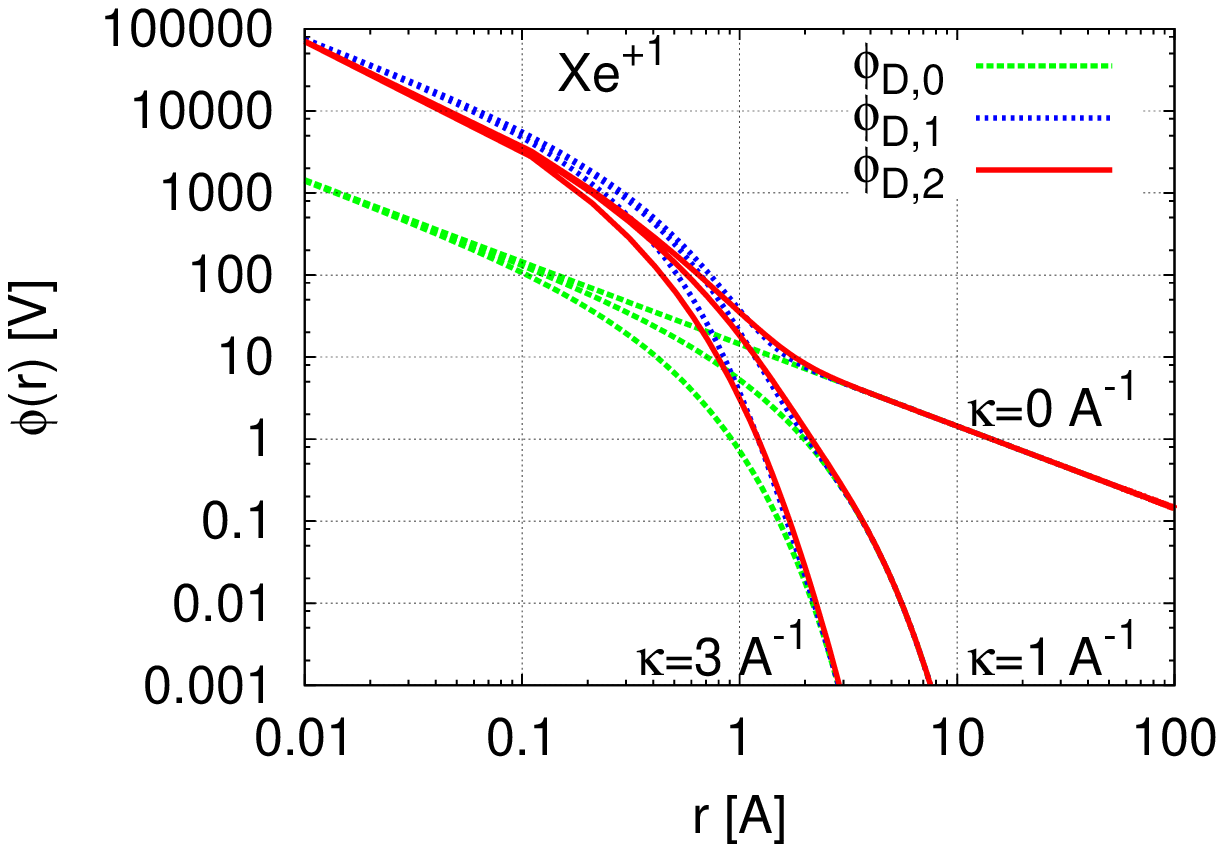}
\epsfig{width=8cm, file=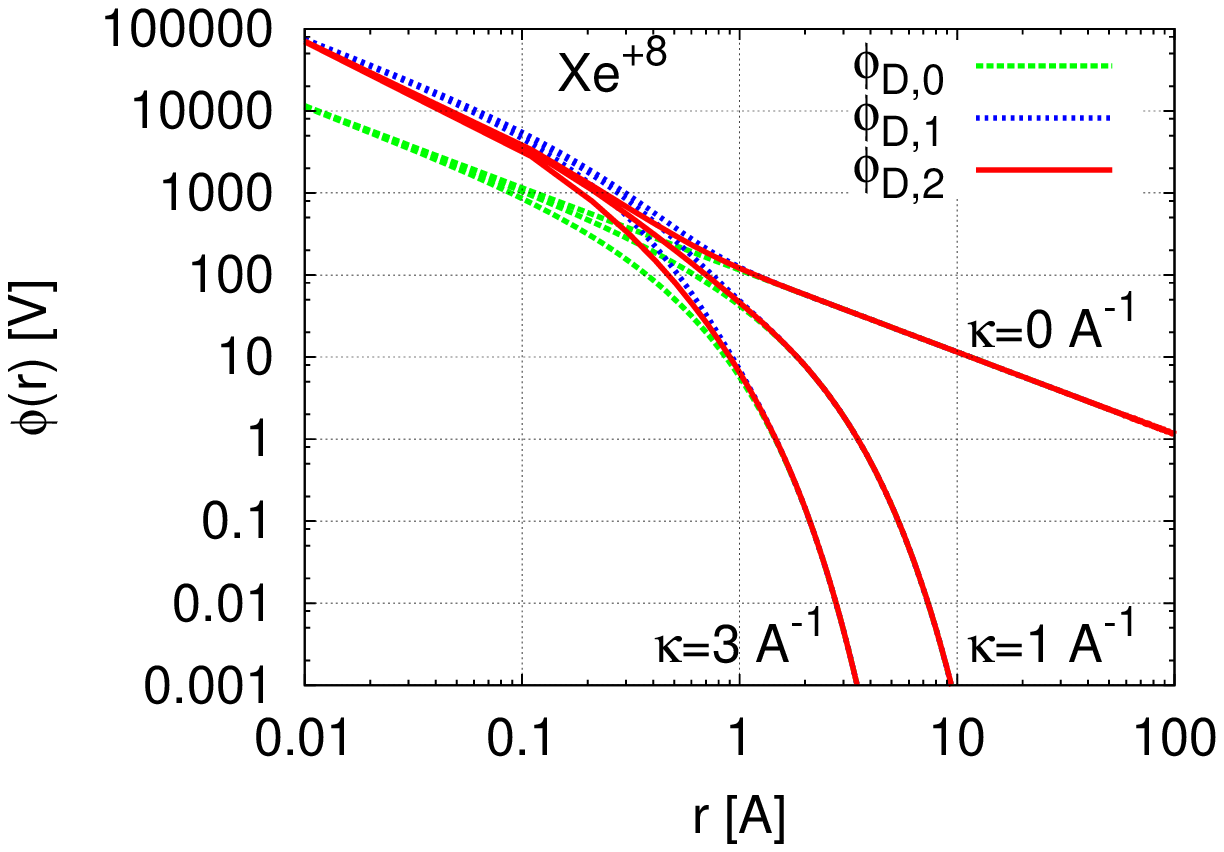}}

\caption{Screened effective atomic potentials of Xe ions within weakly coupled plasma. Potential, $\Phi_{D,1}(r)$, from Eq.\ \re{om1} and potential, $\Phi_{D,2}(r)$, from Eq.\ \re{om2} are plotted for $Xe^{+1}$ ion (left), and $Xe^{+8}$ ion (right) at three different values of $\kappa\equiv \lambda_D^{-1}$, $\kappa=0,1,3$ \AA$^{-1}$. The corresponding Debye screened point-like potentials, $\Phi_{D,0}(r)$, are also plotted for comparison.}
\label{dpotentials}
\end{figure}

We will first consider the classical plasma regime, where the classical statistical mechanics can be applied to model the screening. If the Coulomb interaction within plasma is weak, $\Gamma \ll 1$, mean field estimates for charge densities can be linearized and the Poisson equation for the potential reduces  to the Helmholtz equation of the form:
\eq
(\nabla^2 - \kappa^2)\phi_D(r)=-4 \pi \el \rho(r)\,{1\over\ 4\pi\epsilon_0 }.
\label{helm}
\eqx
Solution of this equation can be obtained by the convolution of the charge density with the Green function,
$G({\bf r})=e^{-\kappa \mid{\bf r} \mid }/\mid {\bf r} \mid$. 
At $\kappa=0$ this Green function reduces to the Green function for the unscreened Coulomb potential, $G({\bf r})=1/\mid {\bf r} \mid$. The general solution of the Helmholtz equation
then reads:
\eq
\phi_D(r)={1\over\ 4\pi\epsilon_0 }\,\int\,d^3 r^{\prime}\, 
{e^{-\kappa \mid{\bf r-r^{\prime}} \mid } \over \mid {\bf r-r^{\prime}}\mid}
\,\el\,\rho(r^{\prime})
\label{solu}
\eqx
\noindent
If the source density is spherically symmetric, the integral over the spherical angle can be performed. If the potential is investigated far away from the source, the dipole approximation can be applied in the expansion of the term,
${e^{-\kappa \mid{\bf r-r^{\prime}} \mid } \over \mid {\bf r-r^{\prime}}\mid}$. Eq.\ \re{solu} then reduces to:
\eq
\phi_D(r)={1\over\ 4\pi\epsilon_0}\,
\int\,dr^{\prime}\,{r^{\prime}}^2\,4\pi\,\el \rho(r^{\prime})
\left( \theta(r-r^{\prime})\,{e^{-\kappa r}\over r} +
\theta(r^{\prime}-r)\,{e^{-\kappa r^{\prime}}\over r^{\prime}} \right),
\label{solred}
\eqx
where $\theta(r^{\prime})$ is the step function.
For the effective atomic potentials defined in the previous section, the weakly screened potential reads:
\eq
\phi_{D,1}(r)={1\over\ 4\pi\epsilon_0 }\,\left( 
{{i \el}\over{r}}\,e^{-\kappa r} 
+{{(Z-i)\el \alpha_i \kappa }\over{\kappa+\alpha_i}}
\,e^{-(\kappa+\alpha_i)r}
+{{(Z-i)\el }\over{r}}\,e^{-(\kappa+\alpha_i)r}
 \right),
\label{om1}
\eqx
for $\Omega_1(r)=e^{-\alpha_i r}$, and: 
{\footnotesize
\eq
\phi_{D,2}(r)={1\over\ 4\pi\epsilon_0 }\,\left( 
{{i \el}\over{r}}\,e^{-\kappa r}
+(Z-i)\el \left( e^{-\kappa r} 
\left[ {{\Omega_2(r)}\over r }+\kappa \Omega_2(r) \right]
-\kappa^2\,\int_{r}^{\infty}\,dr^{\prime}\,e^{-\kappa r^{\prime}}\,
\Omega_2(r^{\prime}) \right) \right),
\label{om2}
\eqx
}
for $\Omega_2(r)=\left[ {\eta \over \zeta}( e^{\zeta r}-1 ) +1 \right]^{-1}$.
We note that in the limit, $\kappa \ll 1$ \AA$^{-1}$, the term,\\ ${{(Z-i)\el \alpha_i \kappa }\over{\kappa+\alpha_i}}\,e^{-(\kappa+\alpha_i)r}$ in \re{om1} 
can be neglected, and the potential, $\phi_{D,1}$, approaches the approximate screened potential used in Ref.\ \cite{santra}.
These potentials are plotted in Fig.\ \ref{dpotentials} for two Xe ionization states: $Xe^{+1}$ and $Xe^{+8}$, at three different values of $\kappa=0,1,3$ \AA$^{-1}$. Despite the different parametrization of the density of bound electrons, the two screened effective potentials are close to each other.  The largest discrepancy between the effective and the point-like potentials occurs at distances less or comparable with the atomic size. Asymptotic limits of all potentials are identical.

The results obtained so far depend on the assumption of weak coupling permitting the linearization of the Poisson equation. If any of the plasma species
is strongly coupled, this approximation is no longer valid and another approach
should be applied. The potentials near the target ions can then be computed by putting each ion into a separate cell. The electrons are divided between the cells in order to provide net charge neutrality to each cell \cite{murillo}. The electron density around the target ion can be approximated as uniform. The Poisson equation for each cell then reads:
\eq
\nabla^2 \phi_{IS}(r)=-4 \pi \el \rho(r)+ 4 \pi \el \rho_e,
\label{is}
\eqx
where $\rho_e$ is the uniform density of free electrons in this cell, and $\rho(r)$ is the ion density \re{gest} that includes the density of bound electrons. This approximation is called the ion-sphere (IS) model.

\begin{figure}
\vspace*{0.5cm}
\centerline{\epsfig{width=8cm, file=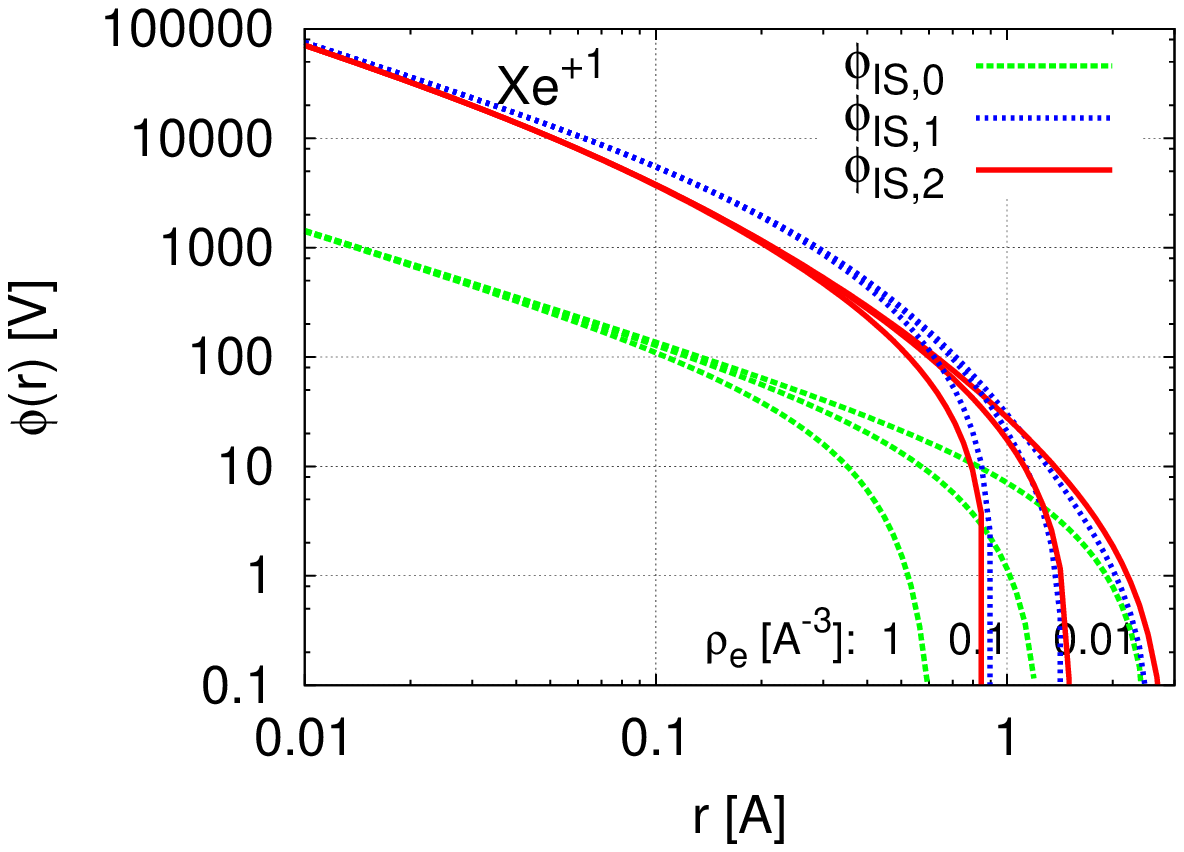}
\epsfig{width=8cm, file=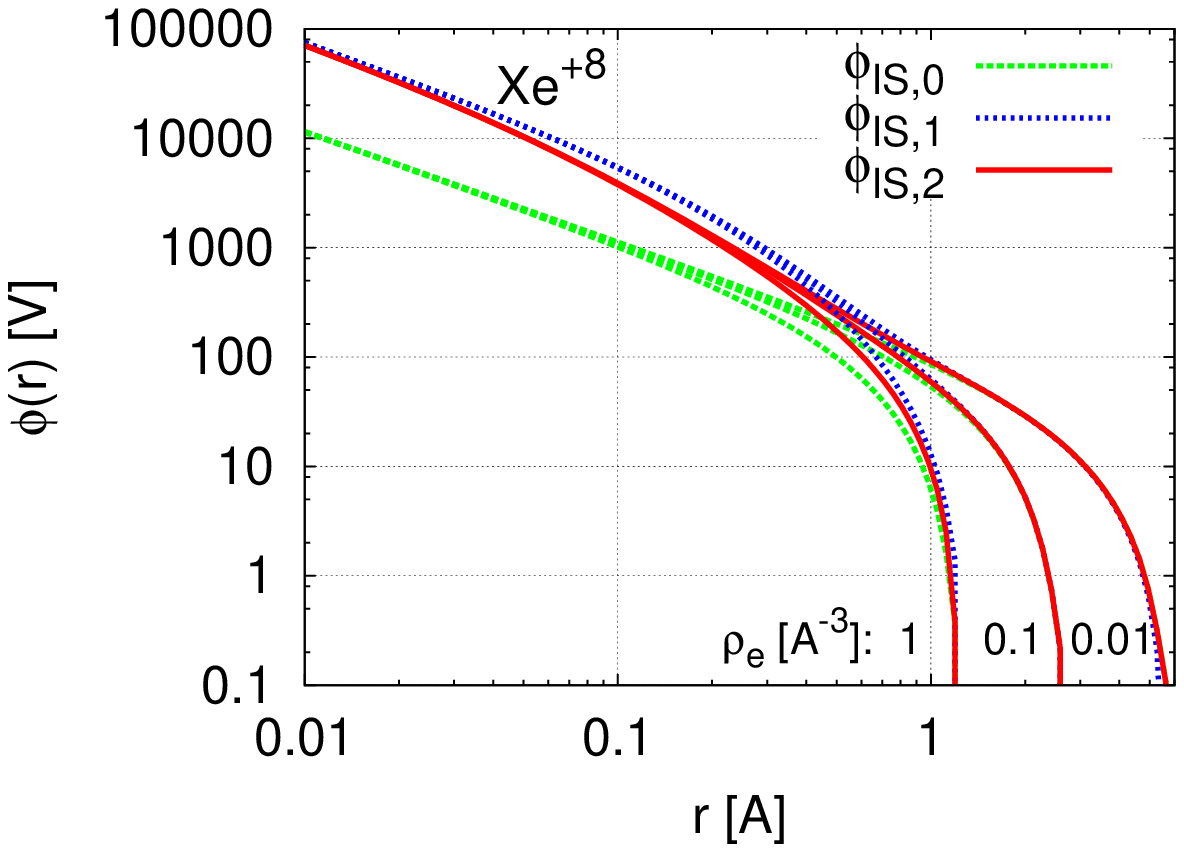}}

\caption{Screened effective atomic potentials of Xe ions within strongly coupled plasma. Ion-sphere approximation is used to model the strong screening. Potential, $\Phi_{IS,1}(r)$, and potential, $\Phi_{IS,2}(r)$, 
from Eq.\ \re{pis} are plotted for $Xe^{+1}$ ion (left), and $Xe^{+8}$ ion (right) at three different values of density of electrons within a cell $\rho_e$, $\rho_e=0.01,0.1,1$ \AA$^{-3}$ (correspondingly $\rho_e=10^{22-24}$ cm$^{-3}$). The IS screened point-like potentials, $\Phi_0(r)$, are also plotted for comparison.}
\label{ispotentials}
\end{figure}

The general solution of this equation is:
\eq
\phi_{IS}(r)=\theta(R-r)\,
\left(\phi(r)- 4 \pi \el \rho_e
\left[{R^2 \over 2}-{r^2 \over 6}\right]\,{1\over\ 4\pi\epsilon_0 }\right)\,
\label{pis}
\eqx
where $\phi(r)$ is the unscreened atomic potential. The radius of ion-sphere cell, $R$, is estimated from the neutrality condition at the cell edge, $\phi_{IS}(R)=0$. For a point-like ion, ${{i \delta(r) }\over{4\pi r^2}}$, the cell size is $R=r_s$, where $r_s=\left( (3 i \el)/(4 \pi \rho_e) \right)^{1/3}$.
For the effective potentials $R$ also depends on the parameters of the unscreened potentials, i.\ e.\ $\alpha_i$, $\zeta$, $\eta$. There is no simple analytical solution of the neutrality condition in this case. The radius of  
a cell can then be either estimated with the asymptotic conditions of this equation at $r=0$ and $r=\infty$, or evaluated numerically. In Fig.\ \ref{ispotentials} we plot the screened effective potentials within strongly coupled plasma at  three different values of the density of free electrons, $\rho_e=0.01, 0.1, 1$ \AA$^{-3}$, within a cell for $Xe^{+1}$ and $Xe^{+8}$ ions. As in case of weak coupling, the screened effective potentials are close to each other. At $r<1$ \AA$\,\,$ the largest discrepancy between the effective and the point-like potentials occurs. However, the application of the effective potentials extends the size of the cell, when compared to the case with point-like ions. This effect is more pronounced for Xe$^{+1}$ ion, and less for highly charged $Xe^{+8}$ ion, when the contribution of the point-like term in the effective potentials dominates, and sizes of cells are nearly identical for all potentials considered.


\section{Cross sections for elastic electron-ion scattering within plasma}

The quantum mechanical cross section for the elastic scattering of an electron
on a central potential in first Born approximation is obtained from the corresponding scattering amplitude, that is proportional to the Fourier transform of the scattering potential \cite{joa}:
\eq
\kphi(\Delta k)={-1\over \Delta k}\,\int_0^{\infty}\,dr\,r\,sin(\Delta k\cdot r)\,\phi(r).
\label{ft}
\eqx
where ${\bf \Delta k}={\bf k_i}-{\bf k_f}$ is the wave vector transfer, $k_i,k_f$ are the wave vectors of the electron before and after the collision.
The differential cross section then reads:
\eq
{{d \sigma_{EL,B}}\over {d \Omega}}=\mid \kphi(\Delta k) \mid^2\,
\left( 2 m \over \hbar^2 \right)^2,
\label{sigel}
\eqx
where $m$ is electron mass.
For the screened effective potentials as defined in the previous section,
the scattering amplitudes in first Born approximation are calculated as:
{\footnotesize
\eqn
\kphi_{D,1}(\dk)&=&-{1\over\ 4\pi\epsilon_0 }\left[ 
{i \el \over (\dk)^2+\kappa^2}
+2\,{(Z-i) \el \alpha \kappa \over ((\dk)^2+(\kappa+\alpha)^2)^2}
+{(Z-i) \el \over (\dk)^2+(\kappa+\alpha)^2}
\right]\label{d1}\\
\kphi_{D,2}(\dk)&=&-{1\over\ 4\pi\epsilon_0 }\left[
{i \el \over (\dk)^2+\kappa^2}+
{(Z-i) \el \over \dk} 
\,\int_0^{\infty}\,dr\,sin(\dk \cdot r)\,e^{-\kappa r}\,\Omega_2(r) 
\nonumber \right. \\
&&\rule{1.2cm}{0cm}\,+{(Z-i) \el \kappa \over \dk}
\,\int_0^{\infty}\,dr\,r\,sin(\dk \cdot r)\,e^{-\kappa r}\,
\Omega_2(r)\,\nonumber \\
&&\rule{1.2cm}{0cm}\,-\left.{(Z-i) \el \kappa^2 \over \dk}
\,\int_0^{\infty}\,dr\,\left(
{sin(\dk \cdot r) \over (\dk)^2}
-{r\,cos(\dk \cdot r) \over \dk} 
\right)\,e^{-\kappa r}\,\Omega_2(r)\,
\right]\label{d2}
\eqnx
}
for the weak coupling case, and:
{\footnotesize
\eqn
\kphi_{IS,l}(\dk)&=&-{1\over\ 4\pi\epsilon_0 }\left[
{i \el \over (\dk)^2}\left[1-cos(\dk\cdot R)\right] 
+{(Z-i) \el \over \dk}
\,\int_0^{R}\,dr\,sin(\dk \cdot r)\, \Omega_l(r)\,\right.\nonumber\\
&&\rule{1.2cm}{0cm}\,-\left. (4 \pi \el \rho_e) \left( 
-{R^3 \over 3}\,{cos(\dk \cdot R) \over \dk^2} + {sin(\dk \cdot R) \over (\dk)^5}
-{R\,cos(\dk \cdot R) \over (\dk)^4} 
\right)\right]
\label{is12}
\eqnx
}
for the strong coupling case, where indices, $l=1,2$ refer to the potential from Ref.\ \cite{santra} and to the potential from Ref.\ \cite{garvey}, respectively. Below we summarize our results on the elastic cross sections obtained for $Xe^{+1}$ and $Xe^{+8}$ (plots not shown).
The cross sections calculated with effective ion potentials in the weak coupling regime change extensively with the parameter, $\kappa$: they decrease with  increasing $\kappa$.
For comparison, at $\Delta k=0.1$ \AA$^{-1}$ the ratio of the cross sections,
$R_D\equiv {d\sigma(\kappa=0)/d\Omega \over d\sigma(\kappa=10)/d\Omega}$ is
$R_D \sim 10^5$ for Xe$^{+1}$ and $R_D\sim 10^7$ for Xe$^{+8}$.
In contrast, the cross sections in the strong coupling approach change less
at the considered electron densities, $\rho_e=0.01-1$ \AA$^{-3}$. They decrease with the increasing density, and $R_{IS} \equiv {d\sigma(\rho_e=0.01)/d\Omega \over d\sigma(\rho_e=1)/d\Omega}$ is $R_{IS}\sim 5$ for Xe$^{+1}$, and 
$R_{IS}\sim 10^2$ for Xe$^{+8}$, also at $\Delta k=0.1$ \AA$^{-1}$. 

For point-like ions the impact of both weak and strong screening effect on the elastic cross sections is much larger: i) for both Xe$^{+1}$ and Xe$^{+8}$
ions, $R_D=10^8$, in the weak coupling regime, and ii) $R_{IS}=40$ for both  Xe$^{+1,+8}$ in the strong coupling regime. 


\section{Cross sections for stimulated photoemission and photoabsorption}


If an electron scatters on an ion in the presence of an external laser field,
absorption or emission of radiation quanta may occur. The quantum mechanical cross section for this process was derived by Kroll and Watson in Ref.\ \cite{kroll}. It sums the individual cross sections for the exchange of $n$ radiation photons:
\eq
\left( {{d \sigma}\over {d \Omega^{\prime}}}\right)_{IB}=\sum_{n=-\infty,n \neq 0}^{\infty}
{{d \sigma_n}\over {d \Omega^{\prime}}}
=\sum_{n=-\infty,n \neq 0}^{\infty}\,{v_0^{\prime} \over v_0}\,
J_n^2(s\,\dv\, cos(\Theta_{\dv,\epsilon}) )\,
{{d \sigma_{EL,B}}\over {d \Omega^{\prime}}}(\Delta v)
\label{kro}
\eqx
where $v_0$ and $E_0$ denote time averaged velocity and kinetic energy of the 
incoming electron, $v_0^{\prime}$ and $E_0^{\prime}$ denote velocity 
and kinetic energy of the outcoming electron and $\Delta v$ is the magnitude of the velocity transfer, $\Delta v=\mid {\bf \Delta v}\mid$. The kinetic energies of the incoming and outcoming electron fulfill the relation: 
$E_0^{\prime}=E_0+n\,\hbar \omega$, where $\hbar \omega$ is the photon energy. 
The field strength parameter, $s$, is defined as $s={ {e{\bf E}_0}\over {\hbar \omega^2} }$. The angle $\Theta_{\dv,\epsilon}$ measures the angle between the vector, ${\bf \Delta v}$, and the vector of the field polarization, $\bf \epsilon$.

Equation \re{kro} holds whenever the Born approximation provides an accurate
description of the elastic process, in which case the elastic cross section
depends on the velocity transfer only \cite{kroll}. We note here that the velocity transfer in this case is due not only to the change of electron momentum but also to the change of the velocity magnitude after emission or absorption of radiation photons. Asymptotics of the Bessel function, $J_n$, implies that at small values of argument, $s\,\dv \ll 1$ (low field and/or slow electrons) single photon exchanges dominate. At high values
of $s\,\dv \gg 1$, the envelope of Bessel function behaves like $(s\,\dv)^{-1/2}$. This yields the classical limit of stimulated absorption and emission cross section (the impact model) \cite{rozmus}.

\begin{figure}
\vspace*{0.5cm}
\centerline{a)\epsfig{width=7cm, file=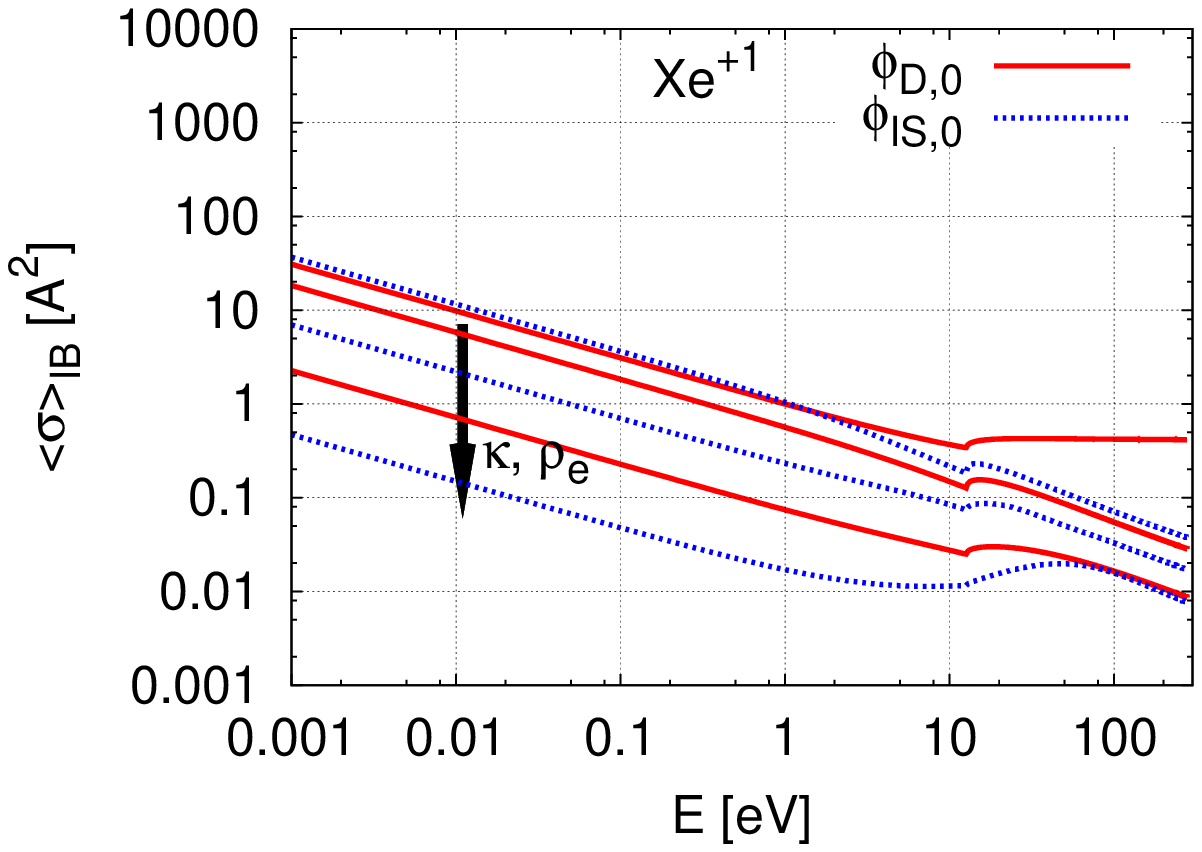}
b)\epsfig{width=7cm, file=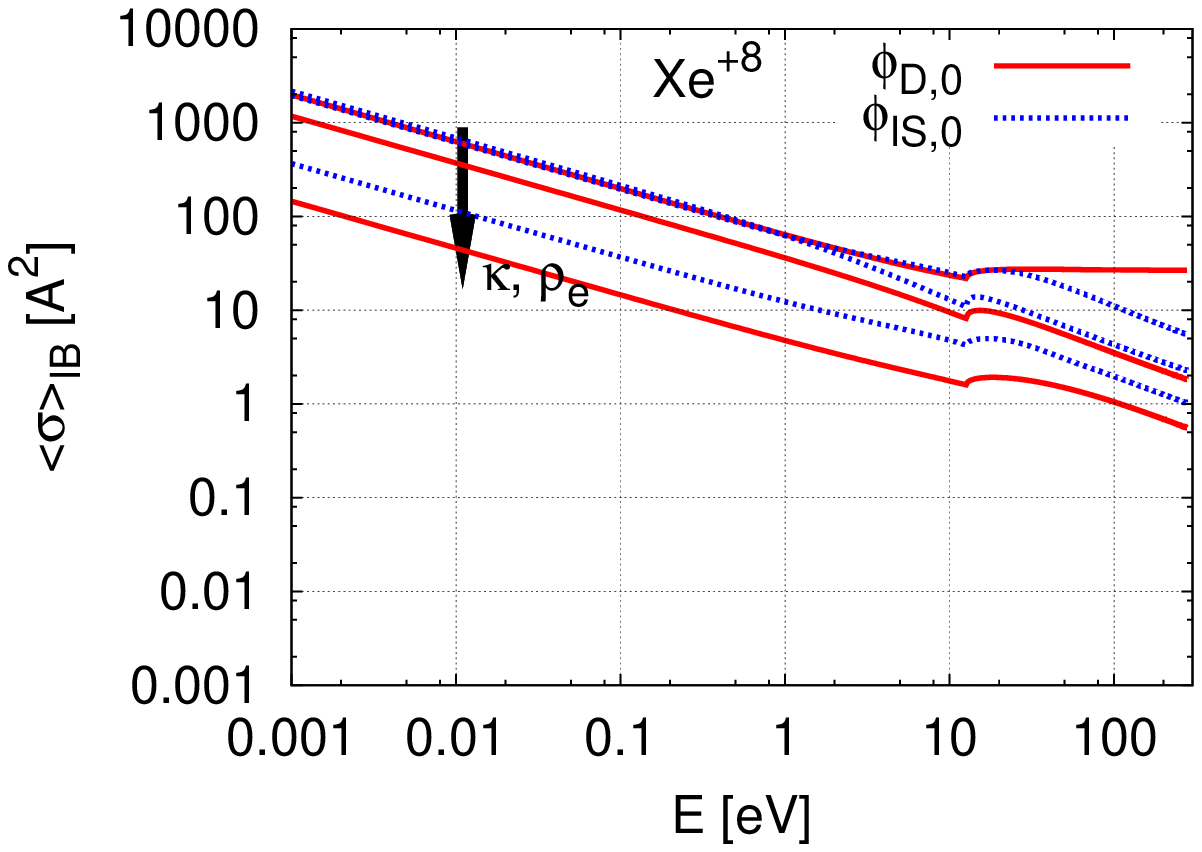} }
\vspace{0.5cm}

\centerline{c)\epsfig{width=7cm, file=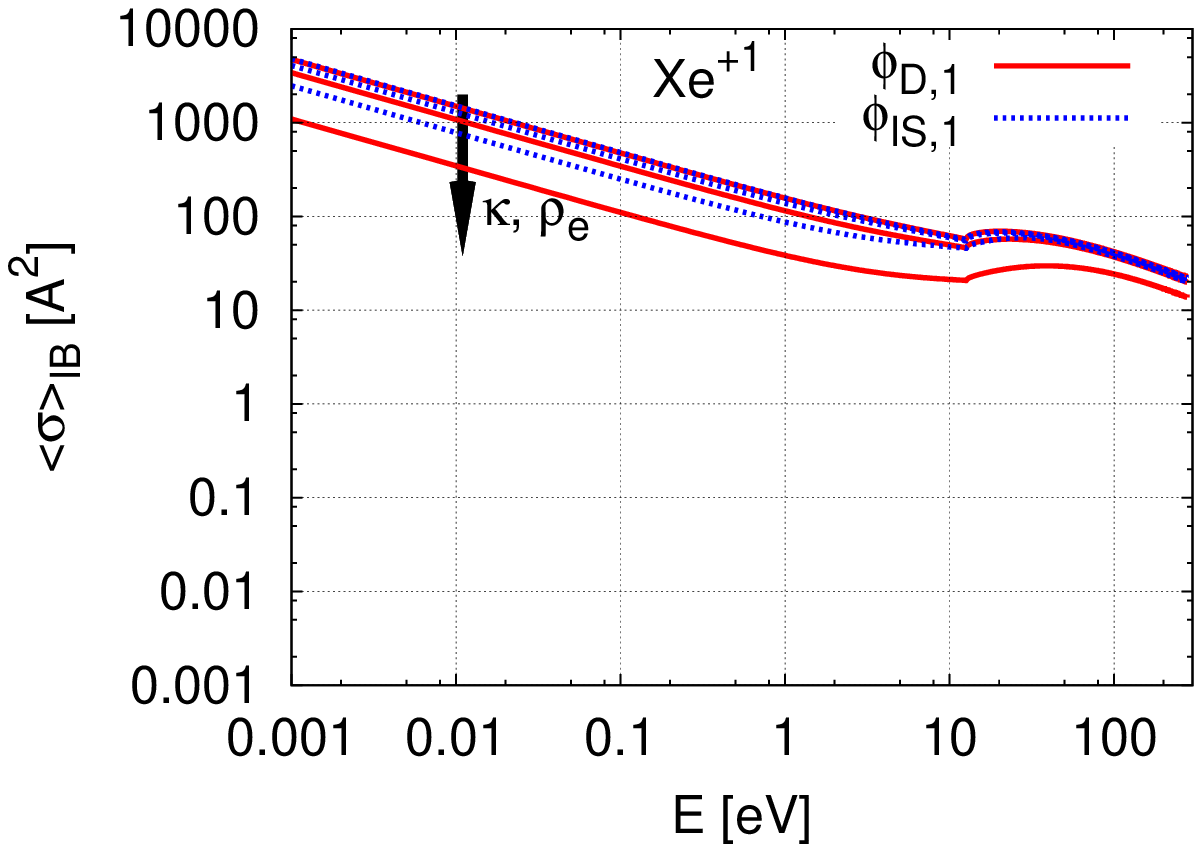}
d)\epsfig{width=7cm, file=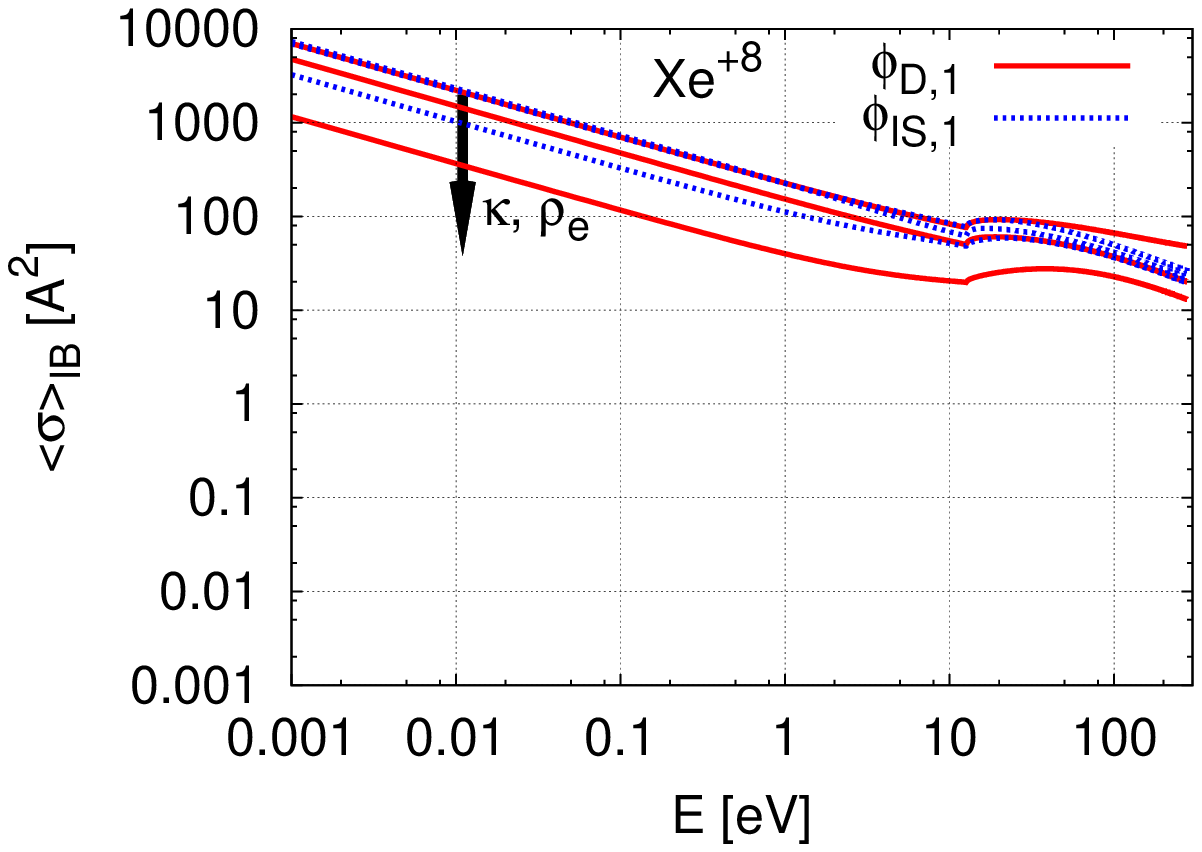} }
\vspace{0.5cm}

\centerline{e)\epsfig{width=7cm, file=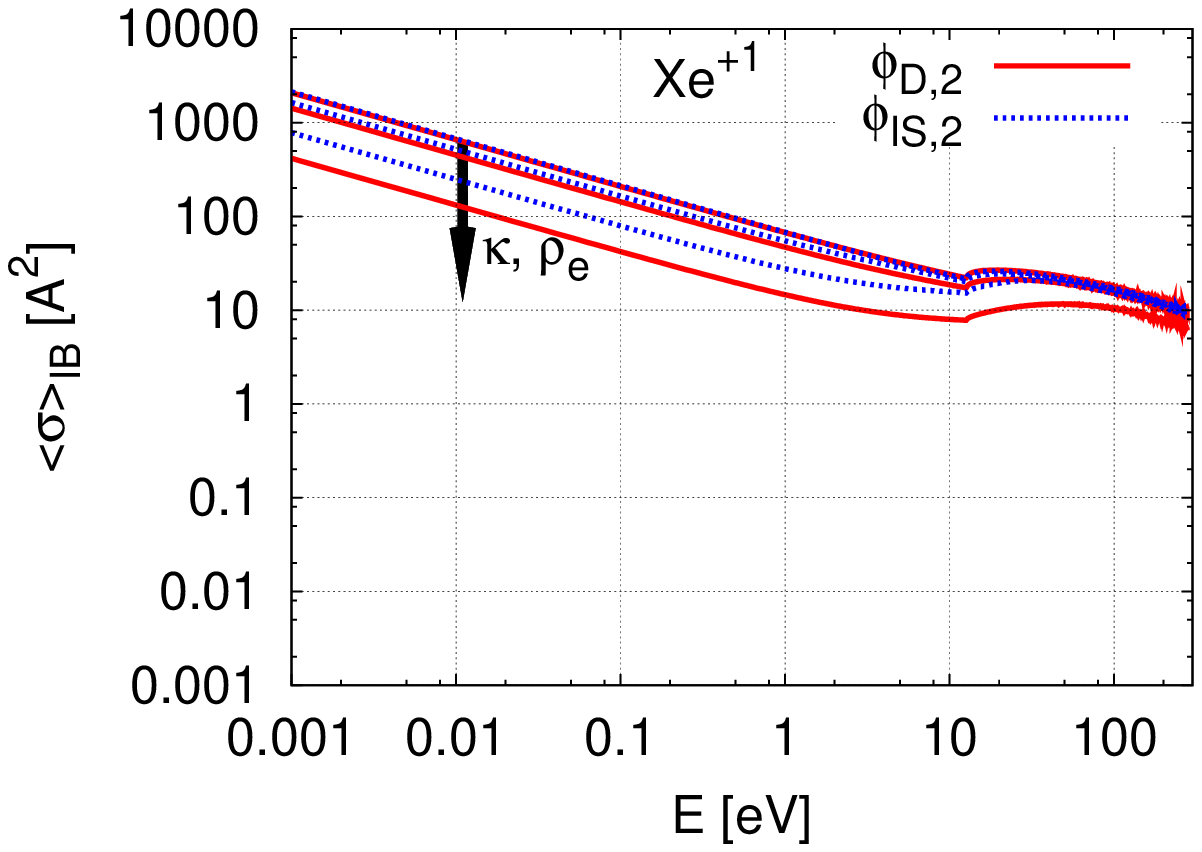}
f)\epsfig{width=7cm, file=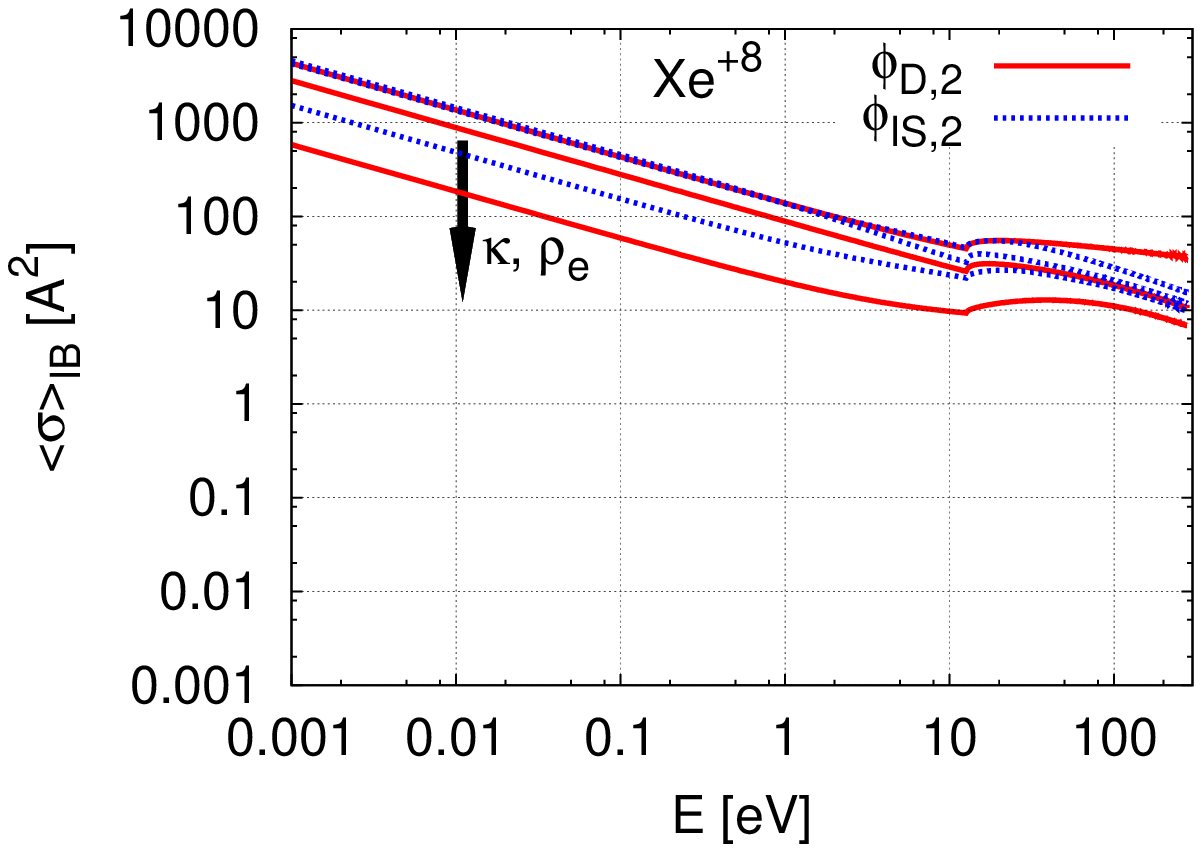} }
\vspace{0.5cm}

\caption{Averaged cross sections for stimulated photoemission and photoabsorption obtained for various atomic potentials: (a-b) point-like Coulomb potential, (c-d) effective atomic potential from Ref.\ \cite{santra}, (e-f) effective atomic potential from Ref.\ \cite{garvey}, at the fixed value of field strength parameter, $s=0.01$ fs/\AA. Results for Xe$^{+1}$ ions (left) and Xe$^{+8}$ ions (right) were obtained at three different values of $\kappa=0,1,3$ \AA$^{-1}$ (weak screening case) and $\rho_e=0.01,0.1,1$
\AA$^{-3}$ (strong screening case) and are plotted as function of the kinetic energy of the incoming electron. Arrows show how the cross sections change
with the increasing values of $\kappa$ and $\rho_e$.}
\label{full}
\end{figure}

We integrate Eq.\ \re{kro} over the scattering angle and average over the direction of the field polarization, $\bf \epsilon$:
\eq
\langle \sigma \rangle_{IB} = {1 \over {4 \pi}}\,
\int d\Omega_{\epsilon}\,d\Omega^{\prime}\,
{{d \sigma}\over {d \Omega^{\prime}}}.
\label{kro1}
\eqx
Below we plot the averaged total cross section for stimulated photoemission
and photoabsorption as a function of the kinetic energy of the incoming electron for the various parametrizations of the atomic potentials of $Xe^{+1}$ and $Xe^{+8}$ ions (Fig.\ \ref{full}). The value of the field strength parameter, $s$, was chosen to match the experimental conditions during the first cluster experiment, $s=0.01$ fs/\AA$\,\,$ at $I\leq 10^{14}$ W/cm$^2$ and at the photon energy, $E_{\gamma}=12.7$ eV. 

For both parametrizations of the effective atomic potentials we obtain similar values of the total IB cross sections, $\langle \sigma \rangle_{IB}$. These values are much higher than the corresponding ones obtained for the point-like potential, i.e. about $150$ times larger for $Xe^{+1}$ ions and $4$ times larger for $Xe^{+8}$ ions. 
The discrepancy of the cross sections is smaller in case of highly charged ion,
as the contribution of the point-like term to the effective potentials Eqs.\ 
\re{d1}, \re{d2},\re{is12} is then much larger than in case of singly charged ions. A significant increase of the cross sections estimated with effective potentials in respect to the cross section estimated with point like potentials has been first observed in Refs. \cite{santra,deiss}, and has lead to the hypothesis of the enhanced plasma heating.

As next we characterize how the plasma environment affects the total IB cross section. As our results should estimate these cross sections within a changing plasma environment, e.\ g.\ in case of transition from
the strongly coupled to the weakly coupled plasma regime, we again consider a broad range of plasma parameters: i) $\kappa=0,1,3$ \AA$^{-1}$ in weakly coupled regime, ii) $\rho_e=0.01,0.1,1$ \AA$^{-3}$ ($\rho_e=10^{22}, 10^{23},10^{24}$ cm$^{-3}$).

For point-like potentials the total cross sections are strongly affected by the plasma environment. At the considered plasma parameters the ratio of the maximal and the minimal cross sections, $R\equiv{\sigma_{max} \over \sigma_{min} }$, is $R\leq 60$ for Xe$^{+1}$ ion and $R \leq 12$ for Xe$^{+8}$ ion. The corresponding ratios estimated, using the effective potentials are: 
i) $R \leq 4 $ for Xe$^{+1}$ and  $R \leq 6 $ for Xe$^{+8}$ with the parametrization from Ref.\ \cite{santra}, and ii) $R \leq 5 $ for Xe$^{+1}$ 
and  $R \leq 7 $ for Xe$^{+8}$ with the parametrization from 
Ref.\ \cite{garvey}. The maximal expected change of the cross sections obtained
using effective potentials can then be estimated with a factor of $7$.

\section{Summary}
To sum up, we have calculated the total cross section for stimulated photoabsorption and photoemission using point-like and effective atomic
potentials within an evolving plasma. The effect of a possible transition from the strongly coupled to the weakly coupled regime on the cross section was evaluated. 

The application of the effective atomic potentials increased significantly the total IB cross sections by a factor of $150$ for Xe$^{+1}$ ions and by a factor $4$ for Xe$^{+8}$ ions in respect to the corresponding cross sections calculated with point-like potentials. Similar effect was observed in Refs. \cite{santra,santra1}.

The total cross sections for photoemissions and absorptions obtained with effective atomic potentials can change by a factor of 7 at most for
plasma parameters in the range: $\kappa=0,1,3$ \AA$^{-1}$ and $\rho_e=0.01,0.1,1$\AA$^{-3}$, and at a fixed value of the field strength parameter, $s=0.01$ fs/\AA. This range of the plasma parameters correspond to that one expected for plasmas created during the interaction of intense VUV radiation from a FEL with xenon clusters \cite{desy}. In contrast, for point-like ions the maximal change of the cross sections is much larger and may be of the order of $60$. 

Our results show that the inverse bremsstrahlung cross sections estimated with 
effective atomic potentials are not much affected by the changing plasma environment. This observation validates the estimate of the enhanced plasma heating effect from \cite{santra,santra1} that may explain the high energy absorption within clusters irradiated with intense VUV radiation.

\section*{Acknowledgements}

Beata Ziaja is grateful to Cornelia Deiss, Wojciech Rozmus and Robin Santra for illuminating comments. 


\end{document}